\documentclass[final,1p,times]{elsarticle} 
\usepackage{graphicx} 
\usepackage{amssymb} 
\usepackage{amsthm} 
\usepackage{lineno} 

\usepackage{bm}

\journal{Nuclear Physics A} 
\begin{document} 

\begin{frontmatter} 


\title{Can the energy dependence of elliptic flow reveal the QGP phase 
transition?}

\author{Ulrich Heinz}

\address{Department of Physics, The Ohio State University, Columbus,
         OH  43210, USA}

\begin{abstract} 
Ideal hydrodynamic simulations are performed to compute the evolution
with collision energy of hadron spectra and elliptic flow between AGS and
LHC energies. We argue that viscous effects should decrease with increasing 
energy, improving the applicability of ideal fluid dynamics at higher
energies. We show that the increasing radial flow at higher energies
pushes the elliptic flow to larger transverse momenta, leading to
a peaking and subsequent decrease of the {\em elliptic flow at fixed $p_T$}
with increasing collision energy, independent of whether or not there
is a phase transition in the equation of state.
\end{abstract} 

\end{frontmatter} 



\section{Introduction and summary}
%
Ideal hydrodynamic simulations of the expansion stage of the hot and dense
fireballs created in relativistic heavy-ion collisions predict a 
non-monotonic collision energy dependence of the ($p_T$-integrated) 
elliptic flow $v_2(p_T)$ \cite{Kolb:1999it}. The softening of the
equation of state (EOS) at the quark-hadron phase transition leads to
a predicted reduction of $v_2$ at RHIC energies, down from SPS energies,
followed by another increase towards LHC energies. This effect is not
seen in experiment \cite{Alt:2003ab} which shows instead a monotonic 
increase of $v_2$ with $\sqrt{s}$. This is now understood as a failure
of the ideal fluid picture during the late hadron gas stage which is
highly viscous and inhibits the buildup of elliptic flow 
\cite{TeaneyPhD,Hirano:2005xf}. Both viscous hydrodynamics \cite{Song:2008si}
and hydro+cascade hybrid algorithms \cite{TeaneyPhD,Hirano:2005xf} reproduce
qualitatively the experimentally observed monotonic beam energy dependence
of the integrated elliptic flow. Viscosity, in particular its strong increase
in the hadronic phase, thus washes out the phase transition transition 
signature in the integrated elliptic flow excitation function.

The PHENIX Collaboration observed that the \emph{$p_T$-differential} elliptic
flow $v_2(p_T)$, on the other hand, when plotted at fixed $p_T$ as a function
of $\sqrt{s}$, shows signs of saturation at RHIC energies \cite{PHENIXv2}.
This has been interpreted as a possible remnant of the non-monotonic energy
dependence predicted by hydrodynamics, signalling the softening of the EOS
near $T_c$ and, possibly, even the existence of a critical end point (CEP) in
the QCD phase diagram \cite{Lacey:2006bc}. The apparent contradiction between
a monotonically rising $v_2(\sqrt{s})$ as observed by NA49 and STAR
\cite{Alt:2003ab} and a saturation withn increasing $\sqrt{s}$ of the 
differential elliptic flow $v_2(p_T)$ is resolved by the observation that
also the {\em radial} flow increases monotonically with $\sqrt{s}$, leading
to flatter $p_T$-spectra at higher energies and thus pushing the 
hydrodynamically generated momentum anisotropy, which is reflected in
$v_2(p_T)$, to larger transverse momenta.  
 
Within a hydrodynamic picture of the collision fireball's collective 
evolution, the monotonic increase with $\sqrt{s}$ of radial flow is a simple 
and unavoidable consequence of energy conservation, independent of (and
at most tempered by \cite{Kolb:1999it}) the existence of a phase transition 
in the QCD phase diagram. A systematic analysis of the hadron $p_T$-spectra
and $v_2(p_T)$ as functions of collision energy \cite{Kestin:2008bh} shows
that a non-monotonic $\sqrt{s}$-dependence of the elliptic flow $v_2(p_T)$
at fixed $p_T$, first rising from AGS to low SPS energies but then falling 
again towards RHIC and the LHC, is a generic consequence of the evolution of 
radial flow and, as such, cannot be used unambiguously as evidence in support 
or against the existence of the quark-hadron phase transition. To make this 
point is the purpose of this contribution. When searching for a clear QCD 
phase transition signature (in particular for the CEP), one has to look 
elsewhere.

\section{Ideal fluid dynamics from RHIC to LHC}

The analysis of Ref.~\cite{Kestin:2008bh} which is reported here is based
on ideal relativistic fluid dynamics (IRFD). As already discussed above,
IRFD is not perfect at RHIC energies and becomes increasingly worse at lower 
energies, due to the growing dynamical role played by the highly viscous 
hadron gas stage. At higher energies, the role of the hadronic phase 
decreases since more and more of the finally observed collective flow (in 
particular its anisotropy in non-central collisions) is generated already 
during the quark-gluon plasma (QGP) stage. The specific shear viscosity 
$\eta/s$ of the QGP (where $s$ is its entropy density) is known to be very 
small, of the order of at most a few times the KSS \cite{Kovtun:2004de} 
bound $\eta/s=1/4\pi$ \cite{Luzum:2008cw}. For fixed $\eta/s$, viscous
effects in heavy-ion collisions are largest at early times, due to the large
initial expansion rate from approximately boost-invariant longitudinal
expansion. At any given early time $\tau$ (before the onset of significant 
transverse expansion, $\tau \ll R/c_s$, where $R$ is the transverse fireball
radius and $c_s$ is the sound speed), viscous effects are controlled by
the ratio of times scales $\frac{\Gamma_s}{\tau}=\frac{\eta}{s}\,\frac{1}
{T\tau}$, where $\Gamma_s=\eta/(sT)$ is the sound attenuation length and
$1/\tau$ is the longitudinal expansion rate \cite{Teaney:2003kp}. In 
perturbative QCD, the dimensionless specific shear viscosity $\eta/s$ is
expected to increase only logarithmically with $T$ \cite{QCD-Viscosity}.
Hence, $\Gamma_s$ is expected to decrease, leading (at the same $\tau$) to 
smaller viscous effects on hydrodynamic flow. Correspondingly, the 
validity of the IRFD approach should improve from RHIC to LHC.

To extrapolate from lower to higher collision energies, we assume that
thermalization occurs earlier at higher densities, i.e. at constant product
$T_0\tau_0=\mathrm{const.}$ Using entropy conservation in IRFD, we can
relate the final charged multiplicity to $T_0$ and $\tau_0$ as follows:
$dN_\mathrm{ch}/dy \sim dS/dy = \tau_0\int d^2x_\perp\,s(\bm{x}_\perp,\tau_0)
\sim s_0 \tau_0 \sim \tau_0 T_0^3$ where $s_0 \sim T_0^3$ is the peak 
value of the entropy density at $\tau_0$ in central collisions. Combining
both conditions we see that, starting from well-established initial
conditions for 200\,$A$\,GeV Au+Au collisions at RHIC \cite{Kestin:2008bh},
the initial thermalization time $\tau_0$ and peak entropy density $s_0$
scale as $\tau_0\sim\left(\frac{dN_\mathrm{ch}}{dy}\right)^{-1/2}$,
$s_0\sim\left(\frac{dN_\mathrm{ch}}{dy}\right)^{3/2}$. The value of
$\frac{dN_\mathrm{ch}}{dy}$ for Pb+Pb at LHC energies cannot be predicted
by hydrodynamics, but will be measured on the first day of LHC Pb-beam
operation. We therefore present our results as a function of 
$\frac{dN_\mathrm{ch}}{dy}$ or, equivalently, of $s_0$. In 
\cite{Kestin:2008bh}, the range $s_0\leq 270$\,fm$^{-3}$
($\frac{dN_\mathrm{ch}}{dy}\leq 1200$) was explored; central $200\,A$\,GeV 
Au+Au collisions at RHIC correspond to $s_0=117$\,fm$^{-3}$ and
$\frac{dN_\mathrm{ch}}{dy}=685$.

\section{Results}

\begin{figure}[t]
\includegraphics[width=0.323\linewidth,clip=]{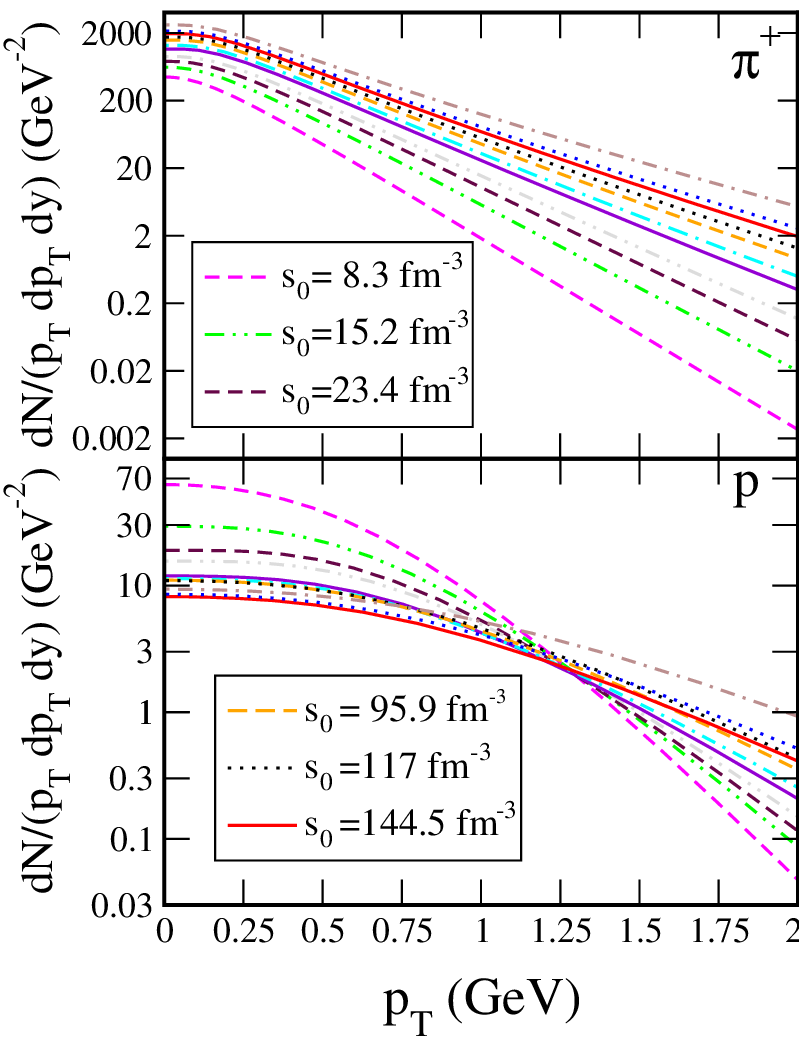} \hspace*{1mm}
\includegraphics[width=0.315\linewidth,clip=]{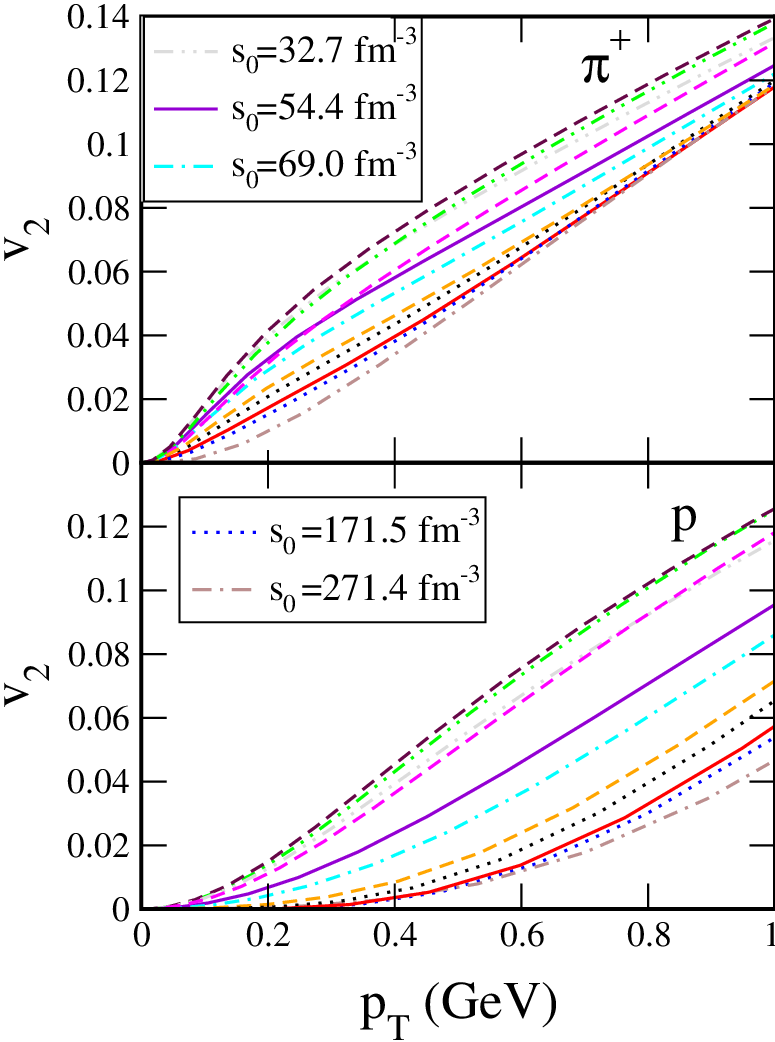} \hspace*{1mm}
\includegraphics[width=0.325\linewidth,clip=]{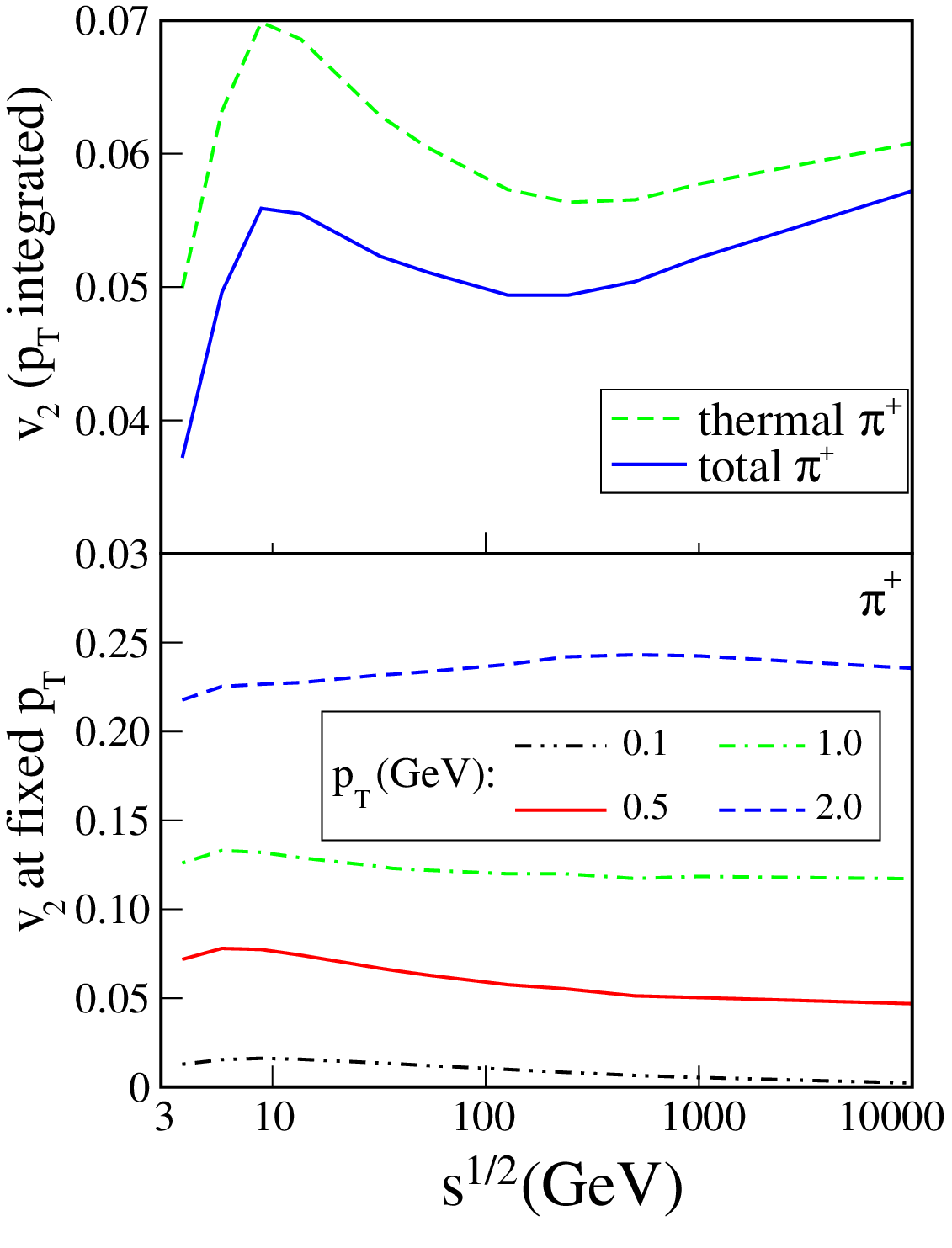}
\vspace{-0.5cm} \caption{(Color online) {\sl Left:} Evolution of pion (top) 
and proton (bottom) transverse momentum spectra for central Au+Au collisions 
from low AGS to LHC energies (to correlate $s_0$ values with collision 
energies and charged hadron multiplicities, see Fig.~1 in 
\cite{Kestin:2008bh}). {\sl Middle:} Evolution of pion (top) and proton 
(bottom) differential elliptic flow in non-central Au+Au collisions 
at $b=7$\,fm. {\sl Right:} Pion elliptic flow as function of collision 
energy (see text for discussion).}
\vspace{0.2cm} \label{Fig1}
\end{figure}
%

The left panel of Fig.~\ref{Fig1} shows the $p_T$-spectra of thermally 
emitted pions and protons (resonance decay contributions not included)
as they evolve from low AGS to LHC energies. The flattening effects of
increasing radial flow are clearly visible, especially for protons where
strengthening radial flow leads to a yield {\em reduction} at low $p_T$ in
spite of the increasing total proton multiplicity. The consequences of
this shape change in the spectra for the $p_T$-differential elliptic
flow is seen in the middle panel of Fig.~\ref{Fig1}, again for thermally
emitted pions (top) and protons (bottom) only. Initially, the differential
elliptic flow $v_2(p_T)$ increases from $s_0=8.3$\,fm$^{-3}$ ($\sqrt{s}
\approx 4$\,GeV) to $s_0=23.4$\,fm$^{-3}$ ($\sqrt{s}\approx 10$\,GeV),
due the increase in total fireball lifetime before freeze-out which
allows more elliptic flow to develop. At higher energies, however,
increasing radial flow pushes the $v_2(p_T)$ curves to the right
(more so for protons than for the lighter pions), leading to a {\em
decrease} of elliptic flow at fixed $p_T$. The bottom part of the right 
panel of Fig.~\ref{Fig1} shows that (for $\sqrt{s}\geq10$\,GeV) this 
decrease of $v_2^{\mathrm{fixed\ }p_T}(\!\sqrt{s})$ is monotonic, and that 
it holds for all $p_T$ values in the range $p_T \leq 1$\,GeV. Plotted
logarithmically, the slope of this decrease is steeper for protons than 
for pions (not shown in Fig.~\ref{Fig1}), reflecting the stronger radial 
flow effects on the heavier protons. 

The radial flow induced decrease of $v_2^{\mathrm{fixed\ }p_T}(\!\sqrt{s})$ is 
independent of the behavior of the $p_T$-integrated elliptic flow, shown in 
the upper part of the right panel of Fig.~\ref{Fig1} for thermally emitted 
(dashed) and all pions (including resonance decays, solid). The integrated 
elliptic flow shows the well-documented non-monotonic behavior of IRFD 
\cite{Kolb:1999it}, featuring a decrease between top AGS and RHIC energies 
caused by the softening EOS near the quark-hadron phase transition, followed 
by an increase above $\sqrt{s} > 100$\,GeV caused by the stiffening of the 
EOS in the QGP phase. The bottom panel shows that, at fixed $p_T$, the 
differential elliptic flow continues to decrease while the 
integrated $v_2$ increases; these tendencies persist to the highest
values of $\sqrt{s}$ where it is known that the elliptic flow fully
saturates in the QGP phase, and that its finally observed value is 
therefore insensitive to the QCD phase transition and to the details of 
the conversion of quarks and gluons to hadrons. In this $\sqrt{s}$-region, 
it is obvious that the decrease of $v_2^{\mathrm{fixed\ }p_T}(\!\sqrt{s})$ is 
unrelated to the softening of the EOS near $T_c$, and has therefore nothing 
at all to do with the phase transition. 

\section{Conclusions}

Energy conservation and hydrodynamic behavior during the fireball
expansion stage lead to increased radial flow from RHIC to LHC and
correspondingly to flatter $p_T$- and $m_T$-spectra, especially for
heavy hadron species. As shown in Ref.~\cite{Kestin:2008bh} this
causes baryon/meson ratios to continue to increase with both $p_T$ 
and $m_T{-}m_0$ at LHC energies, as they do at RHIC. The slope of this
increase as a function of transverse kinetic energy $m_T{-}m_0$ is
almost the same at LHC and RHIC, but as a function of $p_T$ the 
baryon/meson ratios increase with smaller slope at LHC than at RHIC,
due to overall flatter $p_T$-spectra. 

In ideal relativistic fluid dynamics (IRFD), the $p_T$-integrated elliptic 
flow of pions and charged hadrons increases about 10-15\% from RHIC to LHC
energies; accounting additionally for viscous effects at RHIC (mostly of 
hadronic origin) that weaken or disappear at the LHC, the corresponding 
increase is about 25\%. At the same time, the differential elliptic flow at 
fixed $p_T$, $v_2^{\mathrm{fixed\ }p_T}(\!\sqrt{s})$, decreases from RHIC to 
LHC. This decrease is driven by an increase in radial flow which pushes the 
momentum anisotropy to larger $p_T$; it does not depend on a phase transition
in the EOS. Combined with the increase of $v_2^{\mathrm{fixed\ }p_T}
(\!\sqrt{s})$ at low $\sqrt{s}<10$\,GeV, this leads to a non-monotonic
$\sqrt{s}$-dependence of $v_2^{\mathrm{fixed\ }p_T}$ that is {\em generic},
caused by the interplay between radial flow and freeze-out, and not
unambiguously associated with a phase transition in the QCD EOS. Although
the analysis presented here was based on IRFD, the interplay between radial
flow and freeze-out is a general principle that controls the buildup of 
elliptic flow also in real fluids. The observed non-monotonic energy 
dependence of $v_2^{\mathrm{fixed\ }p_T}$ is therefore robust, and (like 
variations of the EOS) inclusion of viscous effects is expected to only 
change the energy {\em where} $v_2^{\mathrm{fixed\ }p_T}$ peaks, but not the 
fact {\em that} it peaks. The search for QCD phase transition signatures, in 
particular for the predicted critical end point connecting a first order 
transition at high baryon density to a smooth cross-over transition at RHIC, 
cannot be based on this non-monotonic energy dependence of fixed-$p_T$ 
elliptic flow.


\section*{Acknowledgments}
This contribution is based on work done in collaboration with G. Kestin
and previously published in \cite{Kestin:2008bh}. Continued support 
by the U.S. Department of Energy under grant DE-FG02-01ER41190 is gratefully
acknowledged.

\end{document}